# DESIGN ARCHITECTURE-BASED ON WEB SERVER AND APPLICATION CLUSTER IN CLOUD ENVIRONMENT

Gita Shah [1], Annappa[2] and K. C. Shet[3]

[1,2,3] Department of Computer Science & Engineering, National Institute of Technology, Karnataka
Surathkal, India
`geet107@gmail.com`, `annappa@ieee.org`, `kcshet@rediffmail.com`

## ABSTRACT

*Cloud has been a computational and storage solution for many data centric organizations. The problem today those organizations are facing from the cloud is in data searching in an efficient manner. A framework is required to distribute the work of searching and fetching from thousands of computers. The data in HDFS is scattered and needs lots of time to retrieve. The major idea is to design a web server in the map phase using the jetty web server which will give a fast and efficient way of searching data in MapReduce paradigm. For real time processing on Hadoop, a searchable mechanism is implemented in HDFS by creating a multilevel index in web server with multi-level index keys. The web server uses to handle traffic throughput. By web clustering technology we can improve the application performance. To keep the work down, the load balancer should automatically be able to distribute load to the newly added nodes in the server.*

## KEYWORDS

*Compute Cloud, Hadoop, MapReduce, load balancing, Web server.*

## 1. INTRODUCTION

Cloud is emerging as a cost effective solution in the computing world. The data capacity of the cloud has gone up to a zettabyte from gigabyte. Cloud Computing are the virtual pool of computing resources that provides an online pool of resources for computing, where the computing is performed by efficient sharing of resources, which we call as load balancing/ resource sharing. Cloud relies on its resources for handling the application in the local and personal devices.

MapReduce [2] is the heartbeat of Hadoop framework. It is a programming paradigm that allows for huge scalability across hundreds or thousands of servers in a cluster. It is a programming model that is associated with the implementation of processing and generating large data sets. The term MapReduce basically refers to two separate and distinct tasks that Hadoop programs perform [6]. MapReduce also plays an important factor in Cloud computing environment, because it decreases the complexity of the distributed programming and is easy to be developing on large clusters of common machine [12]. The data flow architecture of MapReduce shows the techniques to analyze and produce the search index [12] as shown in figure1.

The main purpose of this work is to develop a technique for a fast and efficient way of searching data in the MapReduce paradigm of the Hadoop Distributed File System. Hadoop is a framework which is specifically designed to process and handle vast amounts of data [10]. It is based on the principle of moving computation to the place of data which is cheaper than moving large data blocks to the place of computation.

The Hadoop [1] MapReduce framework is a master - slave architecture, which has a single master server called *a job tracker* and several slave servers called *task trackers*, one per node in the cluster. The job tracker is the point of interaction between users and the framework. The user submits map and reduce jobs to the job tracker, which puts them in a queue of pending jobs and executes them on a first-come-first-served basis.

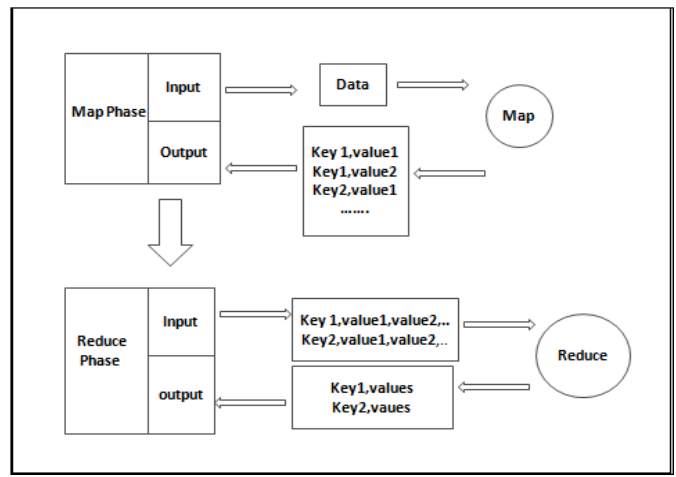

Figure 1. MapReduce data flow

In this paper the schematic view of the design architecture is given, with effective scenario to investigate the performance of Hadoop in high speed retrieval of data in the cloud environment by replacing the map phase of the MapReduce paradigm with a web server. For real time processing on Hadoop, a searchable mechanism is implemented in HDFS by creating multilevel index in web server with multi-level index keys in NameNode.

The rest of the paper is structured as follows. Module 2 & 3 highlights the related works and Hadoop Distributed File System implementation. Module 4 describes the web application in a cloud and architecture design. Module 5 describes the implementation details followed by the results & analysis. Finally work done is concluding, with future work in module 6.

## 2. RELATED WORK

Hadoop [10] is an open source Java-based programming framework which is a part of the Apache project which was sponsored by the Apache Software Foundation. MapReduce is a part of Hadoop framework. It is a programming concept and Hadoop is a framework to run MapReduce program. In February 2003 first MapReduce library was introduced @Google. In 2004 MapReduce was simplified for data processing on large cluster [4]. Google introduced MapReduce in 2004 to support large distributed computing, on large clusters of computers to implement huge amounts of data set [3]. MapReduce is a framework that processed parallel problems of huge data set using a large number of computers or nodes that is collectively referred to as a cluster or a grid [7]. A distributed file system (DFS) facilitates rapid data transfer rates among nodes and allows the system to continue operating uninterrupted in case of a node failure. It was originally developed to support distribution for the Nutch search engine project. In January 2008 Hadoop made apache as a top level project center.

In July 2009 new Hadoop subproject was found where Hadoop core is renamed as Hadoop Common. In this subproject MapReduce and Hadoop distributed file system (HDFS) get separated for doing separate work. HDFS is designed to store very large data sets [10]. In 2010 web application based process is used to handle high throughput traffic [6]. Facebook has claimed that they had the largest Hadoop cluster in the world with 21 PB of storage and on July 27th, 2011 they have announced that the data had grown up to 30PB.

In 2012 the load rebalancing problem in cloud DFSs is illustrated by Hsueh-Yi Chung Che-Wei Chang Hung Chang Hsiao, Yu-Change Chao. The file system in cloud shall incorporate decentralized load rebalancing algorithm to eliminate the performance [17]. In May 2013 a fully distributed load rebalancing algorithm is presented to cope with the load imbalance problem by Hung-Chang, Haiying Shen. This algorithm is a centralized approached in a production system and a competing system [18].

## 3. HADOOP DISTRIBUTED FILE SYSTEM

Recent trends in the analysis of big data sets from scientific applications show adoption of the Google style programming infrastructure, MapReduce [2]. Hadoop, which is a collection of open-source projects originated by "Doug Cutting in 2006" to apply the Google MapReduce programming framework across a distributed system [8]. It provides an easily obtained framework for distributed processing, and a number of open-source projects quickly emerged which leveraged this to solve very specific problems. Hadoop in a cloud computing environment supports the processing of large data sets in a distributed computing environment, primarily in data warehouse systems and plays an important role in support of big data analytics to understand the user behavior and their needs in web services [9].

 HDFS stores file system and application data separately. HDFS also provides high throughput access to the application of data and it is suitable for the application that has large data sets. HDFS is highly fault-tolerant, which provides high throughput access to the application data and is a perfect application that has large data sets. It adopts the master and slave architecture.

HDFS cluster consists of DataNode and NameNode shown in Figure 2. NameNode which is a central server, also called as a Master server, is responsible for managing the file system namespace and client access to files and DataNode in the cluster node are responsible for files storage. In HDFS, a file is divided into one or more number of blocks for storing data [13].

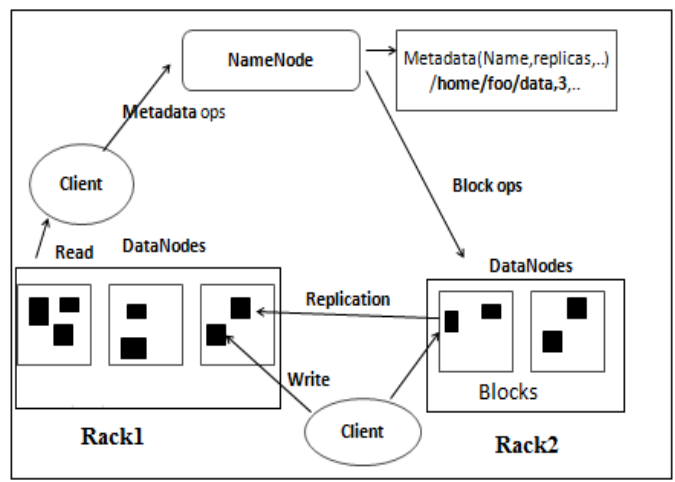

Figure 2. HDFS Architecture

# 3. WEB APPLICATION IN A CLOUD AND ARCHITECTURE DESIGN

In the following module, the schematic design of the architecture for a scaling structure and the dynamic scaling for a web application in cloud is designed. The structure is based on using a front-end load balancer to dynamically route user requests to backend, Master web servers that host the web application. The number of web servers should automatically scale according to the threshold on the number of current active sessions in each web server instance to maintain service quality requirements.

High speed data retrieval from cloud storage has been an active area of research. The Apache Hadoop project design, architecture which is based on the web server and application cluster in cloud environment was started with this as one aim in mind. We review here the basic details of the design architecture, apache load balancer and the details of the Master web server with the multi-level indexing. In the following module, the schematic design of the architecture for a scaling scenario and the dynamic scaling web application for the web in a Cloud is clustered.

## 3.1. ARCHITECTURE DESIGN

The data in Hadoop Distributed File System is scattered, searching and data retrieval is one of the most challenging tasks, so to overcome these tasks a searchable mechanism in HDFS has been implemented. A web server has been designed in the Map phase of MapReduce. The jetty web server is considered as a web interface. Web applications are generated in the HDFS and a web server is started in the form of Hadoop job by submitting our keys to URL Hadoop url/file/key. The Apache load balancer is used in between the client and server to balance the work load of the localhost because if one level is not sufficient, then the system will automatically expand to the second level of indexing.

For real time processing on Hadoop, a searchable mechanism is implemented in HDFS by creating a multilevel index in web server with multi-level index keys. To keep the work down, the load balancer should automatically be able to distribute load to the newly added nodes in the server. The load balancer is used to balance the workload across servers to improve its availability, performance and scalability. To be able to exploit the elasticity of a cloud infrastructure, the applications usually need to be able to scale horizontally, i.e. it must be possible to add and remove nodes offering the same capabilities as the existing ones. In such scenarios, a load balancer is usually used.

NameNode is divided into Master and Slave server. Master server contains multi - level index and slave have data and key index. In the server at the time of the data request, the key is being passed on multi level indexes. Indexing all the process is done by Master server and all retrieval process are done by slave serve. The overall view of the design architecture has been shown in the figure 3.

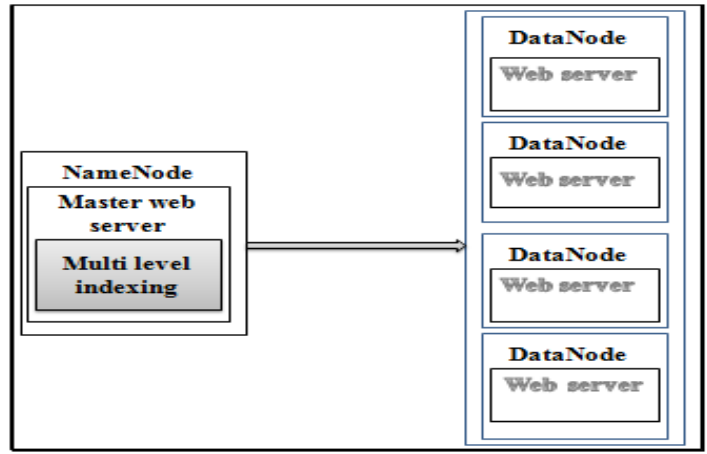

Figure 3. Schematic view of design architecture

### 3.2. LOAD BALANCER

It is used to balance the workload across servers to improve its availability, performance and scalability. Its purpose is to improve the performance of the distributed system through an appropriate distribution of the application [17]. In a distributed system, it is possible for some computers to be heavily loaded while others are lightly loaded. In this situation system can lead to poor system. The goal of load balancing is to improve the performance by balancing the loads among computers.

The Apache load balancer will give better performance of the workload. These can be configured to know a certain number of workers and their host names, but no workers can be added automatically during runtime [14]. It is implemented in between clients and servers to balance the work load of the web server. These servers run on each NameNode and DataNode of which web application package is deployed on the HDFS [13].

### 3.2. LOAD BALANCING WITH THE MASTER WEB SERVER

A common approach is used in combination of the Apache module mod proxy [15] and mod proxy balancer [16]. After adding these modules to the Apache Web Server 2, a virtual host configuration has to be created that configures the load balancer and its balance members. By using the directive ProxySet, the load balancing method can be set to balance either by the number requests, by counting the bytes of the traffic, or by pending requests. By using the Balancer Member directive, a worker with the given URL can be introduced to the load balancer, including how much load it should receive.

The Master server runs on each NameNode and DataNode (Figure 4), web server application package is deployed on the HDFS. The web server in NameNode is divided into Master/slave architecture, according to HDFS storage features, it will store all the keys in the indexing form.

All client requests are first sent to the NameNode through URL Hadoop url/file/key, and Master server in NameNode will decide which the Master server either master/slave on DataNode to respond to the request according to the location information of web files stored in the multi-level indexing on NameNode, and then redirected URL to the node and request the same Uri, finally, a connection can be established between the current Master web server and the browser, and sending files and data between them. To be able to exploit the elasticity of a cloud infrastructure, the applications usually need to be able to scale horizontally, i.e. it must be possible to add and remove nodes offering the same capabilities as the existing ones. In such scenarios, a load balancer is usually used.

It is possible to add the new workers directly to the Master server configuration and starting the load balancer, a large number of virtual folders can be defined in the virtual host configuration. Then, whenever a worker registers at the load balancer, the htaccess rewriter writes a forward rule to the. htaccess file located in htdocs/virtual a request sent to, the URL path route/1 is redirected to the first URL specified in the htaccess file, a request sent to, the URL path route/2 to the second, and so on. Since then htaccess file is evaluated for each request, changing it at runtime allows additions as well as removal of workers. With these measures, it is possible to allow workers to register at the load balancer automatically upon their startup and shutdown, thus bypass the need to manually add or remove workers. Even the rules allow for the changes take effect during runtime. Further modifications of the load factor are possible by collecting monitoring data [11].

A starting load factor can be determined by a performance test. Load balancer knows the potential workload of all workers; it can calculate a load factor for each worker proportional to each worker's potential or monitored workload [5].

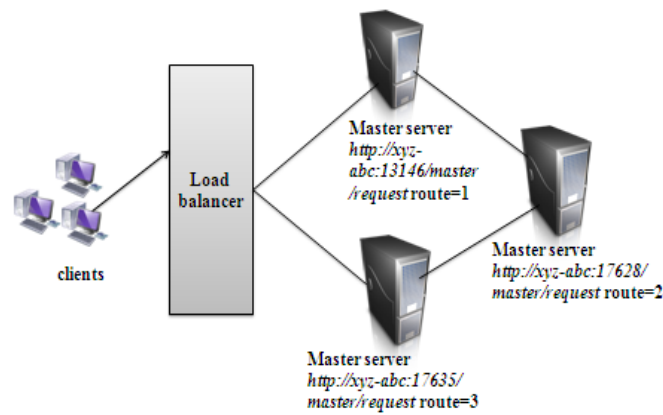

Figure 4. Master web servers based on cluster architecture

The features of load balancer are

- If a server dies connection will be directed to the nearest server.
- Once the server comes back online it rejoins the cluster.
- It can be controlled to take into account machine difference.
- It supports complex topologies and failover configuration
- It distributes the client request across the server, redirect to the localhost.

## 4. RESULT & ANALYSIS

In this module, for the implementation, we configured Hadoop with 6 DataNodes and one NameNode, 6 Task trackers and one job tracker. With each of the nodes we install the jetty server and connect it with Hadoop using eclipse. Jetty server is running on each of DataNode of Master server. Each jetty server works on specific key. Master server contains a list of keys which correspond to jetty server information. When a read request come from the clients via Master server then it does look up to find the jetty server for corresponding keys and then forward it to the corresponding server.  For our experiments we use all systems with Core i7 2.93 GHz CPU, 8 GB DDR3 RAM, 1 TB hard drive. All systems are connected over a 1 Gbps switch.

Web applications are generated in the HDFS and a web server is started in the form of Hadoop job by submitting our keys to the URL of Hadoop i.e. Hadoop url/file/key. The key and value are submitted to the local host of a Hadoop URL in the file system of Master web server and read operations can be performed in the same URL.  The read operation performed in Hadoop is done by Master web server through Uniform Resource Locator. To keep the work down, the load balancer should automatically be able to distribute load to the newly added nodes in the server.

When one Master Server (jetty server) is running and read operation is carried out from HDFS, it has been seen that the average time (seconds) keeps on increasing when data input is increased in HDFS. The average time taken for retrieving data in HDFS is shown in the figure 6 and is 4.2 seconds for 100MB data size, 6.8seconds for 1000MB, 17.3seconds for 10000MB and 24.7second for 100000MB data size.

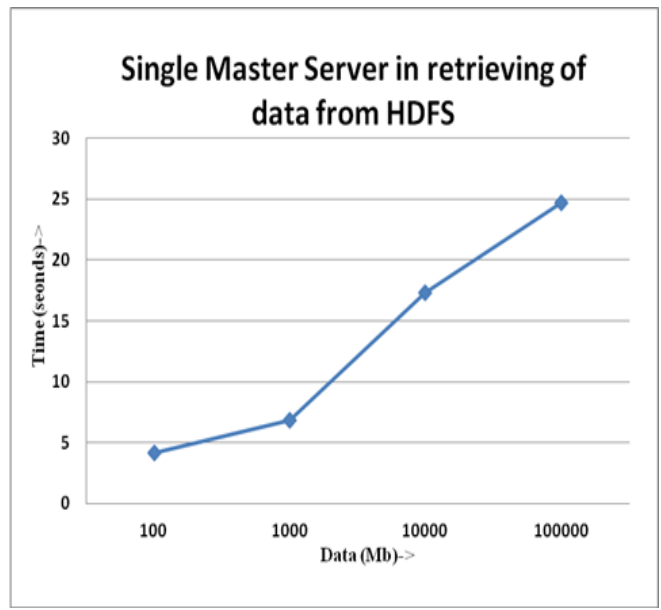

Figure 5. Time taken for single Master server (jetty server) in the retrieval of data from HDFS.

Increased of Master Server (jetty server) will give better performance for retrieving data through the Hadoop url / file/key. It has been seen that the average time (retrieving data) keeps on decreasing when data input is constant or increasing in HDFS.  The Master web server (Jetty server) takes seconds to read a particular or text data from 100000 MB size of data which has been stored in the HDFS where the data are kept constant (100000MB). We have increased the number of servers from 1 to 6 servers for better performance in retrieving of text data from HDFS. The average time taken for retrieval of text data in HDFS by Master server is shown in Figure 7 and is 22.7 seconds, 17.8seconds, and 14.7seconds, 8.3seconds, 7.6seconds and 4.8seconds form 100000MB data size where data are kept constant and web servers are increasing.

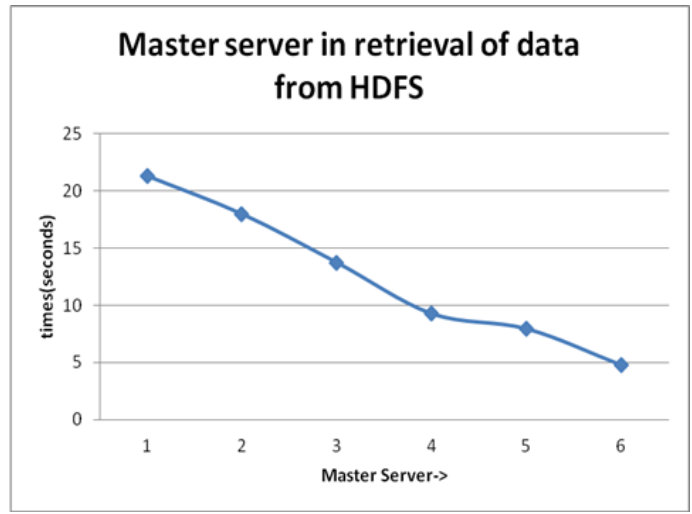

Figure 6. Master Server in retrieval of data from HDFS

## 5. CONCLUSION & FUTURE WORK

In this paper, we have given the schematic view of the designed architecture with effective structure to investigate the performance of Hadoop in high speed retrieval of data in a cloud environment. Apache Web Server a dynamic cloud-ready load balancer was presented as well. It was shown that the Master web server is a static load balancer when using one of the common load balancing modules. To avoid having to manually add workers to the load balancer configuration, we have also added a benchmarking and registration component to the workers.

The results shows that using this architecture in the Hadoop and making map phase as a web server, faster read operation for MapReduce programs can be achieved through URL. For large databases of data warehouse applications in cloud, searching process takes a very long time as the data is scattered. High sped retrieval of data can be particularly useful for real-time applications based on Hadoop where quick fetching and storage of data is necessary.

In future, experiments can be carried out on large datasets and some real-time applications to measure the usefulness of the proposed approach. This web server can conceptually be extended to provide dynamic load balancing capabilities. Apache Web Server cans also be used as a load balancer in combination with Apache Tomcat workers.